# Digital Data Archives as Knowledge Infrastructures: Mediating Data Sharing and Reuse

Revision submitted to the *Journal of the Association for Information Science and Technology*

Christine L. Borgman, UCLA Center for Knowledge Infrastructures, Los Angeles, CA
Christine.Borgman@ucla.edu, https://knowledgeinfrastructures.gseis.ucla.edu/
Andrea Scharnhorst, Data Archiving and Networked Services, The Hague, Netherlands
andrea.scharnhorst@dans.knaw.nl, https://dans.knaw.nl/en
Milena S. Golshan, UCLA Center for Knowledge Infrastructures, Los Angeles, CA
milenagolshan@ucla.edu, https://knowledgeinfrastructures.gseis.ucla.edu/

# ABSTRACT

Digital archives are the preferred means for open access to research data. They play essential roles in knowledge infrastructures – robust networks of people, artifacts, and institutions – but little is known about how they mediate information exchange between stakeholders. We open the "black box" of data archives by studying DANS, the Data Archiving and Networked Services institute of The Netherlands, which manages 50+ years of data from the social sciences, humanities, and other domains. Our interviews, weblogs, ethnography, and document analyses reveal that a few large contributors provide a steady flow of content, but most are academic researchers who submit datasets infrequently and often restrict access to their files. Consumers are a diverse group that overlaps minimally with contributors. Archivists devote about half their time to aiding contributors with curation processes and half to assisting consumers. Given the diversity and infrequency of usage, human assistance in curation and search remains essential. DANS' knowledge infrastructure encompasses public and private stakeholders who contribute, consume, harvest, and serve their data – many of whom did not exist at the time the DANS collections originated – reinforcing the need for continuous investment in digital data archives as their communities, technologies, and services evolve.

# INTRODUCTION AND PROBLEM STATEMENT

Open access to data, or any other profound shift in scholarly practice, does not occur by mandate alone. Rather, change occurs incrementally, as knowledge infrastructures adapt to these new practices. Knowledge infrastructures are "robust networks of people, artifacts, and institutions that generate, share, and maintain specific knowledge about the human and natural worlds" (Edwards, 2010, p. 17). They are living systems influenced by complex sociotechnical factors (Borgman, Darch, et al., 2015; Edwards et al., 2013; Karasti & Blomberg, 2017).

Many stakeholders are involved in the knowledge infrastructures associated with research data. These include the scholars and teams who produce those data, funding agencies that provide the resources to conduct research, universities and other research institutions where investigations are based or conducted, research policymakers in public and private organizations, current and prospective users of those data, and the libraries and archives that may acquire and steward those data. These stakeholders are bound together by community relationships, contracts, and myriad information technologies. Making data "open" occurs in a knowledge infrastructure that mediates exchanges between creators and consumers, both enabling and constraining the uses that can be made of those data.

While research data can be exchanged publicly or privately, data archives are the mechanism preferred by most journals and funding agencies (Borgman, 2015; Pasquetto, 2018; Wallis, Rolando, & Borgman, 2013). Digital data archives play central roles in knowledge infrastructures as entities that facilitate the flow of data between parties, often over long periods of time. Despite the growth in research about data practices, sharing, and reuse, and advances in standards and practices through organizations such as the Research Data Alliance and Force11, few have studied the role of data archives in knowledge infrastructures. All too often, the data archive is a "black box" to which data are contributed and from which data are retrieved (Borgman, Darch, et al., 2015; Borgman, Darch, Sands, Wallis, & Traweek, 2014; Force11, 2018; Mayernik, Wallis, & Borgman, 2013; Mayernik, Wallis, Pepe, & Borgman, 2008; Pasquetto, Randles, & Borgman, 2017; Pasquetto, Sands, & Borgman, 2015; Research Data Alliance, 2018a; Wallis et al., 2013).

The study reported here opens that black box to examine the roles and relationships of data contributors, data consumers, and data curators. Of specific concern are the characteristics and capabilities of knowledge infrastructures supporting data exchange and the mediating roles played by archives as institutions and by archivists as partners with contributors and consumers.

Digital data archives are not monolithic entities; they take many forms and have many homes. Some collect only data of certain types and formats, such as genome sequences for biological research or survey data for the social and economic sciences. Others are more generic, collecting textual documents, static and moving images, audio, and other data types. Data archives range widely in mission, from providing immediate access to replication datasets to long-term preservation. Accordingly, they vary in the degree of investment in data curation. Some institutions devote days or weeks of professional labor to curating each dataset before deposit; others rely on "self-curation," accepting data in whatever form submitted, with minimal review. The longevity of collections also varies from short-term grant funding to long-term commitments by universities, governments, or other agencies (Borgman, 2015; "Directory of Open Access Repositories - SHERPA Services," 2018; International Council of Scientific Unions, 2018; National Science Board (U.S.), 2005). Business models may be based on memberships, grant funding, institutional support, contributions, corporate for-profit entities, or a combination (Shankar, Eschenfelder, & Downey, 2016).

This paper reports on a case study, conducted over a period of three years, of a significant exemplar of digital research data archives: the Dutch Data Archiving and Networked Services institute. DANS was chosen to represent several trends in knowledge infrastructures associated with research data. It serves multiple communities with a diverse array of material spanning the social sciences and humanities, plus some physical and life sciences content. It is a government-funded entity responsible for collecting certain categories of data, thus providing an opportunity to assess the influence of policy mandates. DANS is a node in multiple networks of data repositories and digital libraries, nationally and internationally, being part of intersecting knowledge infrastructures. Lastly, DANS provides "self-archiving" services, holding contributors largely responsible for curation activities prior to deposit. However, they also employ a staff of archivists, providing opportunities to observe the distribution of labor between contributors and staff, and mediation activities with contributors and consumers.

# BACKGROUND AND LITERATURE REVIEW

We briefly review knowledge infrastructures, motivations for sharing and reusing research data, uses and users of digital data archives, and context for our case study.

## Knowledge Infrastructures

Infrastructures are difficult to study because they are complex, long-term, social and technical entities that are largely invisible (Edwards et al., 2013; Karasti & Blomberg, 2017; Star, Bowker, & Neumann, 2003; Star & Ruhleder, 1996). All infrastructures are fragile in the long term, although components such as digital archives may last for decades (Borgman, Darch, Sands, & Golshan, 2016; Edmunds, L'Hours, Rickards, Trilsbeek, & Vardigan, 2016; Lee, Dourish, & Mark, 2006). Some infrastructure components may be field-specific, such as ontologies or genome databases, while others may support multiple domains, such as archives of government statistics or general social surveys. Yet others may be essential resources for multidisciplinary fields, such as the International Ocean Drilling/Discovery Program that provides cores for biological and physical science research (Darch & Borgman, 2016; Darch et al., 2015).

Research data are deeply embedded in the knowledge infrastructures of the research enterprise. Individual researchers and teams conduct their studies with particular sets of instruments and tools, including software and code, and with an underlying theoretical or empirical model. While data are fundamental parts of the research process, extracting them as products to be shared with others can be a fraught endeavor. Such extraction requires extensive labor, expertise, and expense beyond the conduct of the research per se. Incentives and motivations to share and reuse data are complex and differ considerably by domain, funding source, type of data, and other factors. The availability of data for reuse depends on infrastructure to make those data discoverable, retrievable, interpretable, and usable (Borgman, 2015; Bowker, 2005; Edwards et al., 2013; Karasti & Blomberg, 2017; Latour, 1987; Latour & Woolgar, 1986; Star et al., 2003; Star & Ruhleder, 1996).

## Sharing and Reusing Research Data

Research data take many forms and may originate from observations, experiments, excavations, physical specimens, or other methods. Determining what are data is itself problematic, as one person's signal is often another's noise. Here we draw upon Borgman's definition that *data* refers to "entities used as evidence of phenomena for the purposes of research or scholarship" (2015, p. 29). Thus, almost anything can be treated as data in a research project.

Disincentives to release, share, and reuse data often outweigh the incentives for compliance. Incentives among stakeholders also are misaligned, as costs and benefits may be distributed in ways that discourage data sharing. While many researchers recognize the importance of preserving data, others question the long-term value of their data to themselves or to others, or ask whether potential reusers can understand someone else's data (Borgman, 2015; Curty, Crowston, Specht, Grant, & Dalton, 2017; Frank, Yakel, & Faniel, 2015; Gregory, Cousijn, Groth, Scharnhorst, & Wyatt, 2018; Mayernik, 2011, 2016; Piwowar, 2011; Tenopir, Palmer, Metzer, van der Hoeven, & Malone, 2011; Tsoukala et al., 2015; Wallis et al., 2013; Weber, Baker, Thomer, Chao, & Palmer, 2012; Zimmerman, 2008).

Releasing data, whether by sharing directly with other persons or by depositing in a data archive, requires careful selection of the data and work to add metadata and contextual information necessary for interpretation. Software or algorithms associated with data production may be needed to interpret or reuse datasets. The labor involved in documenting data for sharing is often extensive and unrewarded, and requires skills beyond the expertise of most researchers (Borgman, 2015; Mayernik, 2016; Mayernik et al., 2013; Pasquetto et al., 2017, 2015; Wallis et al., 2013). Researchers often maintain a sense of ownership over their data, regardless of legal status. Other reasons to control access to data include protecting privacy, cultural sites, endangered species, and intellectual property rights (Borgman, 2018; Eschenfelder & Johnson, 2014).

## Uses and Users of Digital Data Archives

Despite the long history of studying information-seeking behavior in libraries and other institutional contexts (Case, 2006), uses and users of digital resources have proven harder to investigate. One reason is the multiple roles of those individuals. Researchers may contribute data to archives, may be consumers of data in those archives, or both (Palmer, 2005).

Data sharing and reuse are context-specific, thus archivists wish to understand their communities sufficiently to maintain trusted relationships (Faniel, Barrera-Gomez, Kriesberg, & Yakel, 2013). While professional practice dictates that services, infrastructure, standards, policies, and practices of digital data archives be based on their designated communities of users (Consultative Committee for Space Data Systems, 2012; Yakel, Faniel, Kriesberg, & Yoon, 2013), social science research reveals that communities take many forms and are difficult to characterize (Lave & Wenger, 1991; Wenger, 1998).

More studies have addressed why people search digital data archives than why individuals contribute data (Gregory et al., 2018). As many data archives are field-specific, research tends to focus on searching behavior within specific domains such as archaeology, social sciences, or engineering. Not surprisingly, user behavior tends to correlate with existing data practices in a field, and archives tend to be tailored accordingly. Archaeology, for example, is promoting common standards to improve management of the wide variety of data types and formats in current use (Arbuckle et al., 2014; Faniel, Kansa, Kansa, Barrera-Gomez, & Yakel, 2013; Faniel & Yakel, 2017; Kansa, 2012; Kansa & Kansa, 2011, 2013; Kansa, Kansa, & Arbuckle, 2014).

Assessing the match between content, services, and communities is difficult enough when the users self-identify with a domain. Much harder to study are the archives that serve broad communities with diverse types of data; for example, those that span the social sciences and humanities such as DANS, or institutional repositories that serve all schools and departments of one or more universities. The more generic the data collection, the more diverse the community of users.

## DANS: Data Archiving and Networked Services

As indicated in the problem statement, DANS is an ideal case to examine how digital data archives serve as knowledge infrastructures and how they mediate data sharing and use. DANS was founded in 2005 as an institute of the Royal Netherlands Academy of Arts and Sciences (KNAW) and of the Netherlands Organization for Scientific Research (NWO) (Data Archiving and Networked Services, 2017). However, its archival collection dates back to the 1960s, a growth period for social sciences data archives (Doorn & Tjalsma, 2007). ICPSR in the U.S. (Regents of the University of Michigan, 2016) and the Dutch Steinmetz Archive (whose collections are now part of DANS) ("Steinmetz Archive," 1989) were launched in that time period. DANS serves a broad community, spanning multiple disciplines, and employs a staff of professional archivists with domain expertise. Their commitment to self-archiving, complemented by the curation expertise of the staff, provides an opportunity to examine how dataset documentation and services are mediated.

DANS was among the first data archives to offer web-based data submission. They now provide an array of online services, including NARCIS, the Dutch Research Information System; DATAVERSE.nl, a Harvard-based open-source platform to store, share and register research data; and EASY, the Electronic Self Archiving SYstem. This study focuses on EASY, as an archivist-assisted self-archiving service. An EASY dataset is the equivalent of a "collection" in Dublin Core Metadata Initiative terminology. Datasets are categorized by discipline for purposes of targeting communities for data reuse. The largest number of datasets (75%) originate in archaeology, for which EASY is a legal deposit archive ("EASY: Published datasets," 2016). Most other datasets originate in the social sciences and humanities (Akdag Salah et al., 2012; Scharnhorst, Ten Bosch & Doorn, 2012). Datasets vary widely in size and in number of individual files, which can range from one to several hundred.

This external case study builds upon prior self-studies conducted by DANS. Among the earlier findings are that Dutch researchers' concerns about sharing and reusing data reflect standardization needs, data appraisal and data backlog issues, a sense of ownership over the data, fear of misunderstanding and misinterpretation of their data, and policy requirements (Data Archiving and Networked Services, 2010; Dillo & Doorn, 2011). Quantitative analyses of uploads and downloads of datasets show steady increases over time. More than 85% of datasets in DANS have been downloaded at least once (Doorn, 2017). The most downloaded individual datasets are from the social sciences, such as census data from Statistics Netherlands (Akdag Salah et al., 2012; Scharnhorst, Bosch, & Doorn, 2012). Archaeology represents the largest category of downloads in absolute terms (Doorn, 2017). Increasingly, datasets reach DANS via automatic import of batches of datasets arranged by data librarians in other institutions, rather than individual uploads by human depositors (Dillo & Doorn, 2014).

Initially, DANS services were intended solely for Dutch researchers. As DANS attracted more international users, they became part of a network of Trusted Digital Repositories. These agreements allow DANS to serve as dark archive for other repositories such as DRYAD and the Mendeley Data service. DANS participates in numerous Dutch, European, and international endeavors such as the European data infrastructure for scientific research (EUDAT), Advanced

Research Infrastructure for Archaeological Dataset Networking (ARIADNE), the European Open Science Cloud, and the European Holocaust Research Infrastructure (Advanced Research Infrastructure for Archaeological Dataset Networking, 2012; EUDAT Collaborative Data Infrastructure, 2016; European Holocaust Research Infrastructure, 2015; European Open Science Cloud, 2018). DANS also is involved in efforts to align data sharing policies and practices, such as operationalizing the FAIR principles (Wilkinson et al., 2016), and is a member of international organizations such as ICSU-World Data Systems, Research Data Alliance, Science Europe, and the Digital Preservation Coalition (Digital Preservation Coalition, 2018; International Council of Scientific Unions, 2018; Research Data Alliance, 2018b; Science Europe, 2018).

# RESEARCH QUESTIONS

The case study reported here was developed as part of a larger research agenda to understand how practices vary between research domains and how these factors influence data sharing and reuse ("UCLA Center for Knowledge Infrastructures: Home," 2018). We focus here on the mediating role of digital data archives between contributors and consumers of research data, characteristics of these communities, and their expectations for digital data archiving services. The term "contributor" refers to the person who collected data that were deposited in DANS, regardless of whether the actual uploads of data were done by the contributor or by another person, such as a staff member, librarian, or a graduate student. "Consumer" refers to the person who retrieved or acquired the data from DANS, regardless of whether or how the data were subsequently reused. "Archivists" are staff of DANS who handle datasets (as data managers), interact with users (in communicative functions), or formulate policies and formal requirements to data submission (as technical archivists).

Three research questions guide our study, using DANS as an exemplar of digital data archives:
1. Who contributes data to the digital archive? How, when, and why do they contribute?
2. Who consumes data from the digital archive? How, when, what, and why do they consume?
3. What roles do archivists play in acquiring, curating, and disseminating data?

# RESEARCH METHODS

These research questions and methods were developed in a series of visits to DANS over a three-year period when the first author was a KNAW Visiting Professor, hosted at DANS. Documents about DANS, many of which were in English, and publications by DANS staff and by those interviewed were acquired and analyzed. This study followed the same general methodological framework as our other recent studies of data practices (Borgman, Darch, et al., 2015; Borgman, Golshan, et al., 2016; Darch & Borgman, 2016; Sands, 2017).

An early step in our research was to characterize the communities of contributors and consumers of DANS data, and the degree of overlap, by mining DANS transaction logs of system usage. Like most digital service organizations, they maintain weblogs for purposes of auditing, trouble shooting, and managing information. Users must register with DANS to contribute data or to retrieve certain kinds of datasets, thus creating a user database with a small amount of demographic information (e.g., name, institution, email address, discipline). Although transaction log data have a long history in information retrieval for studying user behavior, they can be difficult to interpret. Logs provide traces of what people do, but lack information about why they do so (Borgman, Hirsh, & Hiller, 1996).

We selected transaction logs and the associated database of registered users from three fiscal years, October 2011 through September 2014 (FY 2012-2014). This was a period of consistent record-keeping since the last major system upgrade. The logs contained sufficient information to identify contributors and consumers, but not frequency of system use (Borgman, Van de Sompel, Scharnhorst, van den Berg, & Treloar, 2015). With a goal of 10 interviews from each group, we drew an initial sample of 50 contributors (from 3517 submissions during the sampling time frame) and 50 consumers (of 3401 registered during the sampling frame). As registration is not required to search DANS, the true number of searchers or consumers cannot be known. After initial low levels of responses, more candidates were randomly selected from the existing pools. In total, we contacted 75 contributors, 9 of whom agreed to participate, and 112 consumers, 8 of whom agreed to participate. Weblogs listed only names and email addresses associated with uploading and downloading files, which was insufficient information to determine whether an individual was the data collector or an intermediary. To reach persons who were associated with data creation and reuse, we requested contacts with researcher-contributors in our interview solicitation. In addition, we interviewed 10 members from the DANS team, which included all of the curators and archivists on staff at the time. Our meetings with DANS senior management were treated as ethnography. In all interactions with DANS staff, we emphasized that our goal was to study DANS as an exemplar of digital data archives, and not to evaluate the organization or the staff.

Most (21 of 27) interviews were conducted in person in the Netherlands, preferably in the offices of the interviewee. These interviews, averaging about one hour in length, were conducted in English by one or two UCLA staff members. In most of the interviews with contributors and consumers, one DANS staff member participated in the interview, providing context and translation (usually of technical terms) as needed. The remaining interviews of contributors and consumers were conducted remotely by videoconference by a UCLA team member. All interviews with DANS staff members were conducted at DANS by a UCLA team member, with minimal participation of DANS staff to maintain confidentiality. While participating DANS interviewers took UCLA training in human subjects research and were certified to participate in the study, all data coding was conducted by UCLA staff and anonymized before sharing with any DANS staff.

Meeting people in the offices and homes where they use resources such as DANS provides context not possible in surveys. These individuals showed us files on their computers, stacks of materials, and representations of artifacts they had created through their use of DANS systems and services. Some walked us through their departments or workspaces, explaining the flows of material in and out of DANS and other information systems. A museum curator gave us a private tour of the current exhibitions, which helped us to understand how digital materials contributed to these physical exhibits. Visiting people in large cities and in small towns also let us see how much, or how little, they relied on colleagues and on digital services for access to information resources.

Open-ended questions allowed us to pursue unanticipated lines of inquiry and to expand on initial responses. Conducting interviews in person is far more time-consuming and expensive than distributing an online survey, thus the rarity of such studies is not surprising. However, given the importance of digital data archives to knowledge infrastructures, scholarly communication, and science policy, such investments in exploratory research are essential (Borgman, 2015; Tenopir et al., 2015; Wallis et al., 2013). While the sample size is small, each interview delivered rich, in-depth information. In total, over 27 hours of interview material were analyzed. Analytical coding of interview transcripts, fieldnotes, and documents were conducted with NVivo, a qualitative analysis software package.

In the findings section below, interview subjects are anonymized, assigned labels based on the category by which they were identified in the weblogs (contributor, consumer) or DANS staff, and given a numerical identifier within the category (e.g., Contributor1).

# FINDINGS

Findings are drawn primarily from the interviews, with ethnography and document analyses serving to frame the study. The contributors represented a diverse array of disciplines, spanning the content of the DANS data collections. The consumers were an even more diverse set of individuals, spanning disciplines and practices well beyond the research community. These two groups overlapped less than anticipated. DANS archivists devoted about half their time to contributors and half to consumers. However, they worked much more intensively with contributors, assisting them in describing and depositing data. They also solicit data for DANS through outreach activities. Archivists' contacts with consumers occurred primarily through the help desk services.

## Sample Demographics

We conducted a total of 27 interviews with 28 people (one interview was conducted with two people from the same organization); 21 men and 7 women. The distribution of research subjects by category, domain, and occupation is presented in Table 1. Domain expertise is based on self-report in the interviews, classification of datasets contributed, and other background material obtained from publications and personal websites. One person, Consumer4, told us in the interview that he also contributed data. He is counted as one consumer interview, but his comments are reported in both categories.

*Table 1: Distribution of Interview Sample*

| Stakeholders/Participants | Number of Interviews | Domain Expertise | Occupation |
|---|---|---|---|
| Data contributors | 9 | Archaeology, history, and related fields (3)<br>Labor economics (1)<br>Linguistics (2)<br>Oral histories (1)<br>Scholarly communication (1)<br>Plant biology (1) | Academic staff (7)<br>Scholarly-professional society (1)<br>Private company employees (2 persons; 1 interview) |
| Data consumers | 8 | Archaeology, history (4)<br>Political science (2)<br>Social science (2) | Academic staff (3)<br>Cultural institution staff (2)<br>Citizen scientists (1)<br>Students (2) |
| DANS staff | 10 | Archaeology and humanities (7)<br>IT development (3) | Archivists, project managers, and IT developers (10) |

## Data in DANS/EASY

DANS acquires data for EASY (Electronic Self-Archiving SYstem) for audiences classified as humanities; social sciences; behavioral and educational sciences; law and public administration; life sciences; medicine and health care; economics and business administration; and interdisciplinary sciences. In the submission process, the producer is asked to assign a dataset to primary and secondary audiences from this classification, which creates a disciplinary index (Scharnhorst et al., 2012). Data types include text, tables, images, graphs, maps, and audio-visual files such as oral history materials. The most common formats are PDF, DOC, and TXT for text; CSV and XLS for tabular data; TIF for images; and MP4 for audiovisual recordings. Archaeology reports usually contain text, tables, photographs, graphs, and maps in a single PDF file (Mientjes, 2015). As of May 2015, when the interview sample was drawn, EASY contained 29,743 published datasets. Datasets average about 100 files each (Doorn, 2017); an example display is shown in Figure 1. Dataset structures may indicate which files are open access and which require permissions. Contributors have considerable flexibility in structuring files, datasets, and granular access control. While this flexibility provides context-specific metadata and organization, it also limits the consistency of data structures within EASY.

*Figure 1: Granularity of Access*

| Name | Size | Accessible |
|---|---|---|
| csv | | All |
| Database design.pdf | 116010 | Yes |
| Dissertation_definitief.pdf | 7616252 | Requires granted permission request |
| Reopenedgraves_Lowcountries.accdb | 8024064 | Yes |

Source: https://easy.dans.knaw.nl/ui/datasets/id/easy-dataset:66658/tab/2

## Contributors of Data to DANS

In most cases, individual researchers contribute data that they collected. This is the simplest situation and the one where interviews provided the fullest explanation of the motivations for contributing, along with the specific practices, processes, and constraints. In other cases, individuals identified as contributors in the log files were intermediaries between researchers and the DANS/EASY system. Some contributors identified other individuals in their organizations who sometimes upload data files to the DANS/EASY system on their behalf, which is another mediating role.

### *Types and Origins of Data*

While the interview sample was too small to achieve proportional representation by research domain, the nine interviewees spanned a wide variety of research areas. Most contributors discussed one or a few datasets they had input to DANS/EASY. The exception was staff from a professional archaeology company, who discussed practices involved in contributing about 250 datasets per year. They are one of about 20 such companies in the Netherlands.

Contributors described their data precisely. Consumer4, the university faculty archaeologist who also contributed data to DANS, distinguished between three types of data based on stages and types of research:

> "The first is excavation data, so field maps, digitized field maps and that sort of information. The second group of information is on finds, lists of finds made during excavations … the third category is reports on individual excavations."

Consumer4 further distinguished between digitized data that were born in analog form and data that were born digital:
> "Until 20 years ago, excavations were drawn at scale on large sheets of paper, and these sheets were stored in an archive…. Nowadays, more and more of these paper drawings have been scanned and are available as bitmaps, rasterized data… (In recent years), … (with) 'robotic total stations,' they make digital measurements in the field, and you have immediately your maps in a digital form."

Contributor5 commented on how heavily his private archaeology company has invested in digital data production. They are testing new data format standards that will automate parts of the ingest process to DANS/EASY and other repositories.

Contributor4, a faculty member, distinguished between types of data collection that acquire physical samples and the ways in which representations are created, processed, and coded. Academic researchers and archaeology companies drill boreholes to assess a field or building sites and take sample cores. As multiple cores may be obtained from each borehole, they are described individually. Site maps that indicate the locations of boreholes are "high level interpreted data or information" to Consumer4. He only shares his maps, created with geographic information systems (GIS), "with the direct workers." While his "basic map is just lines and polygons with labels," his coding system enables the team to run scripts that yield "nicely colored maps…" The files he contributed to DANS/EASY, which are large in number, consist of map layers and parameters that can be used to reconstruct his maps or make new ones.

Contributor7 described the large oral history projects, which include audio recordings of interviews, transcripts, summaries, and extensive metadata, that his team deposited in DANS/EASY for stewardship rather than for access. At least one of his datasets is stored in DANS as a dark archive, not accessible to anyone but the owners, due to political sensitivity. The open materials can be searched and streamed via a project website, delivered in the background by DANS. The audio materials also are being used as a testbed for research in computational linguistics. In this case, DANS' mediating role is a protective one.

*Motivations to Deposit Data*

The majority of the academic researchers with whom we spoke deposit data in DANS/EASY as a means to share them with other researchers within or outside their academic community. As Contributor4 said, "another reason to put it in DANS was I don't want to sit on data until all the publications are ready." Similarly, from Contributor8, "I think it's very important that we share all our data, … the data can be used so many times for so many different questions." Others, such as Contributor3, deposit data for preservation purposes: "I volunteered …when I retired, to make a database of it… So I hope it will never be lost."

Academic researchers also deposit data to fulfill requirements by their funding agencies. Contributor6 was explicit that "the reason we deposit the data there originally has to do with the contractual agreement we have with the people who pay for the data gathering." Legal deposit requirements are explicit for archaeological data. The *European Convention on the Protection of the Archaeological Heritage*, known as the Valletta Treaty, requires archaeological investigations prior to breaking new ground for any construction (Council of Europe, 1992). Netherlands law associated with the treaty requires that reports of these archaeological investigations be made public by deposit in DANS/EASY; some reports also are deposited in local archives and in the national library.

The archaeology company we visited deposits about five reports per week, or about 250 per year. Most of these are initial site surveys. The survey reports we examined average about 90 pages in length; these are PDF documents that contain many tables, maps, and charts, largely in color, with extensive narrative. The entire report is contributed as a single PDF file; internal tables and charts are not structured files. In about 10% of investigations, they find sufficient evidence of human activity such as pot sherds, weapons, or traces of old buildings that a more extensive study is legally required. In about 10% of the latter cases, a full archaeological dig is required, which may postpone building construction for a considerable time period. With a few exceptions, all of these reports are contributed to DANS/EASY.

### *Community Characteristics*

Some, but by no means all, of the contributors of data were also consumers of data. Only two of the academic contributors said that they searched DANS/EASY for data to incorporate in their own research. In contrast, the two individuals from the private archaeology company began each new project by searching DANS for any prior archaeological studies of the region of interest. Their field reports cite prior studies, whether conducted by these or other companies, or academic publications.

Archaeological studies reported in scholarly publications overlap with private company reports in several respects. University and company researchers use each others' data. Academic articles take longer to produce, are more nuanced, and data deposit may follow publication. Private investigations are conducted and filed quickly to fulfill contract requirements and keep construction projects on schedule. These reports tend to follow a standard template.

DANS plays a mediating role in knowledge infrastructures that facilitate data exchange. Some universities in the Netherlands have for-profit units that compete with private companies to conduct archaeological site studies. University faculty also consult for archaeology companies when specialized expertise is required, such as types of pottery, military history, or geomorphology. Archaeology company employees may alert their former professors to interesting new finds. Continuity between generations also occurs when faculty inherit data from their advisors, as we found in another case.

### *Credit, Control, and Attribution*

Data contributors can maintain varying degrees of control over their data, and academic contributors typically released only to registered users. The contributors interviewed appear to assume that consumers interested in their data will register with DANS and make a request to use the data. As Contributor4 said, "they ask, 'Is your data open?' This is not open, open, open; it's open after you announce your name."

Contributors expressed commitment to open access, such as Contributor6: "First of all, we are also believers of the open source, and open data policy, so we try to be as open as we can with the data." Yet they also express concerns: "Sometimes you want to keep some data until you have your stuff published, because otherwise people will publish instead of you, which is bad for us." Contributor6 and his team have an internal policy for archiving the data for their own reuse. Other researchers deposit their data to DANS for preservation and open access, as Contributor3 said, "I hope some people will use it in the future…[and] they will make a publication."

Archaeology datasets are the only subset of DANS for which registration is specifically required, largely due to concerns about protecting sites from looters. To join the DANS/EASY archaeology group, individuals must demonstrate some professional affiliation with the field. Contributors may "lock" their files, making them available only upon request. Access requests go directly to the contributors, who can then negotiate with the requestor. The majority of academic contributors we interviewed preferred to restrict access to users of DANS who have registered by name. They may ignore anonymous requests or ask the requestor to register.

Conditions for providing access to locked data files varied considerably. Contributors were more willing to grant access to data for which they have completed all expected publications. In other cases, they may request a collaborative relationship. Contributor4, for example, was more willing to release data to Dutch researchers than to those from other countries. This was at least partly a matter of convenience as the text was in the Dutch language and others would need translation assistance. In contrast, an academic researcher who ran a citizen science project wanted the data to be available as quickly as possible as a means to keep participants motivated.

Release conditions tend to be project-specific. Consumer4, speaking in his role as a contributor, explained:
> Normally I ask, "What precisely are you looking for, and for what purpose?" Because generally they ask permission to use all the data. And there are maybe 200 datasets within this project, and I can happily give them permission to inspect all these 200 datasets, but if they ask me what they want to know, then I can tell them, "Take a look in these two datasets." Or "Forget it, the data which you need are not there."

Institutional factors also influence release conditions. Contributor3, the retired faculty member, explained how his university had a "front office" relationship with DANS, where the librarians contributed data on behalf of faculty and researchers. Despite his preferences for open access, a librarian had locked his datasets, following institutional practice. As a test, he tried to obtain access to his own data from DANS and was unable to do so. With some investigation, he determined that the request went from DANS to the library staff, who did not realize the library had a responsibility to respond. Next, the request went to a co-author on the original project, who did not respond because he was not involved in the data deposit and unaware of his delegated responsibility. Contributor3 said "in my opinion, … if people want to steal our data without reference to it, let it be. …"

DANS management now encourages contributors to adopt Creative Commons licenses. Several people mentioned receiving the license-change request and their explicit refusal. These individuals would only contribute data to DANS (or other repositories) if they could maintain some degree of access control.

Data citation is a common means to give credit for releasing data. DANS follows archival best practices by assigning a digital object identifier (DOI) and suggesting a citation for each dataset. Despite the ready availability of these metadata, contributors were inconsistent in citing data. Contributor4 always cites his datasets. Contributor6, in contrast, sometimes reports data in a publication without mentioning their availability in DANS/EASY. When individuals requested data from him, he sometimes referred them to DANS and sometimes gave them the data directly.

## Consumers of Data from DANS

The consumers of DANS data were strikingly diverse in terms of discipline, occupation, and the uses to which they put data they acquired. While some were academic researchers, the group also included practitioners and lay persons. They were distributed geographically, in large cities and small towns, from the center to the periphery of the Netherlands. With the exception of Consumer4, who also contributes data to DANS, none of the consumers knew any DANS staff by name.

### *Community Characteristics*

We traveled to the far corners of the Netherlands to meet consumers of DANS data, not only in their offices, but in their homes if they preferred. Most of them used DANS intermittently, at most a few times per year. Several mentioned refreshing their familiarity with the system prior to meeting with us.

Consumer3 contrasted most strongly with the contributors; he was a computer professional whose avocation was guiding local history tours in his small town. He had become an expert on a particular type of regional archaeology, obtaining most of his material from DANS/EASY. He also searched DANS on behalf of his daughter, who was an undergraduate archaeology student. Over tea at his kitchen table, he showed us the database he had constructed and the blog site he maintained for local enthusiasts. A three-ring binder with maps, drawings, and other exhibits encased in plastic sheets was his guidebook for leading tours in the summer. In winter he did his research, and also was friendly with the town's official archaeologists and archivists.

Consumer5, in another small town, was a curator at a regional museum. Our assumption, and his initial explanation, was that he was using DANS to obtain material for his museum exhibits. As the conversation progressed, a broader array of uses for DANS emerged. He also was using DANS/EASY as a digital library because the archaeological reports contained more extensive descriptive content and bibliographies than he could acquire from local libraries or from the collections of his museum. Outside of his professional time, often on evenings and weekends, he searched DANS/EASY for the doctoral research he plans to pursue upon retirement from the museum. He was excited to show us a large collection of three-ring binders in which he had cataloged all known specimens of a certain type of bronze object found in the Netherlands. Each record had an image and extensive metadata of his own devising.

*Information Seeking*

Consumers sought information about particular sites or regions in the Netherlands, or about particular type of objects. The ability to search in the Dutch language for Dutch materials was a significant attraction, as Consumer8, an archaeologist, said, "There's really no other repository available in the Netherlands. There's only DANS… mainly of interest for Dutch resources, and it's the sites, and the raw data, it's only in Dutch..."

Keyword searching in DANS for place names or locations is a daunting task due to the variety of data types and formats, the complexity of naming geographic locations, and the mix of Dutch language with English abstracts. Consumers typically browsed content by category or geographic region, exhibiting varying degrees of sophistication. Consumer8 referred to the "community name" ...where the excavation or the research took place." Consumer4 used the more technical term, to say that "datasets are usually stored under toponyms… if you cannot guess the toponym under which the dataset is stored, then you will not find it. … searching on the municipality, then you get enormous lists of datasets." Consumer3, the local guide, mentioned that he was willing to browse through 200 or so matches, which was four to five screen pages.

Consumer4 was one of several to identify the need for a "mapping facility" by which a searcher "can make geographical selections in the DANS archives." DANS recently had added a capability to search sites by drawing a region on an interactive map of the Netherlands. The new feature was not widely advertised, however, and not readily apparent in the interface. In several cases, we took a few minutes after the end of the interview to demonstrate this new feature, to the great interest of these interviewees.

*Uses of Data*

The uses of DANS/EASY data were also diverse. Consumer3, the local guide, is "mostly interested in maps combined with texts … so I can walk around and say to the people, 'This is... But this and this happened.'" Contributor5, of the commercial archaeology company, searches DANS by region to "make an inventory of the excavations and investigations in the neighborhood of our plan area." Consumer8 uses DANS archaeology reports to make comparisons in the field, "if it contains what I'm looking for, then I … check the field drawings and the photographs (to) see (if) what the archaeologist wrote … is consistent with the way it looks like in the field, through my eye."

Consumer4, found "enormous profits" from DANS for "an assignment... that can only be done by combining very old data with very recent data." He frequently finds both the "very old data … And all the recent data" in DANS/EASY. His willingness to be interviewed was partly a payback for the benefits of access to DANS and its services. Consumer4 also told us that he prefers "raw data, which I can connect in my own way, to very connected data, which are very difficult to entangle … rather 10 datasets with the various components of the data separate, than one big dataset, and I have to split all the data to get access to these parts… useful for my purposes."

DANS data also are used for course assignments, as Consumer1, a graduate student, explained: "We had to download a file …a workbook ...with all kinds of hyperlinks to DANS' archives."

Consumer6, now a faculty member in her own country, had obtained data from DANS while working as a scholar in the Netherlands. She had not yet used those data, several years after acquiring them, but planned to use them in future research.

*Credit, Control, and Attribution*

Contributors who lock their data appear to assume a peer relationship with potential consumers. Consumer4, who is also a contributor, is "prepared to invest some time in helping people who request his data and not just saying, 'Okay. There are 200 datasets and good luck with it.' Because it gives me a moral right to ask for some support when I need data as well, because the dataset may contain... thousands of files."

However, the consumers we interviewed, especially the non-academic ones, were seeking data that are easily accessible. As Consumer3, the local guide said, "when I can't access it, I don't know whether it's interesting. I could ask for access, but … I haven't tried, no…" While he is a registered archaeology user, he was finding enough material already and would rather limit his search to browsing unrestricted data.

**Archival Staff of DANS**

DANS employs archivists, technical staff, administrators, and researchers spanning the domains represented in its collection. They assist contributors with metadata, migration to archival formats, documentation, and ingest processes. Archivists staff help desks and respond to about 1,000 queries per year by email or phone; by rotating help desk responsibility, staff are cross-trained in the many processes, practices, and problems that arise. Experience in working with users also contributes to software design and maintenance. We interviewed the ten individuals directly responsible for DANS/EASY as archivists, managers, software and systems development, and related roles. To maintain anonymity in this small sample, we report all as "DANS staff."

*Knowledge Infrastructure Activities*

The work of DANS staff is best understood in relation to the larger knowledge infrastructure in which DANS/EASY operates. DANS is a government agency responsible for collecting certain kinds of material and to which specific kinds of material must be submitted. To fulfill these responsibilities, DANS has collaborators in the Netherlands, Europe, and elsewhere. Partners include government agencies that contribute census and statistics data, institutions that contract with DANS to contribute other kinds of data, universities and libraries who partner with DANS on "front end" services, agencies who require deposit with DANS, the Netherlands Royal Library, and the Dutch funding agencies that provide continuing support. DANS also provides services and databases associated with their many research projects.

DANS creates metadata for datasets in DANS/EASY, and operates NARCIS, which includes a name authority file for Dutch researchers (Reijnhoudt, Stamper, Börner, Baars, & Scharnhorst, 2012). In turn, metadata in DANS databases are harvested by other services such as Europeana, OCLC, and Google Scholar. They also provide some identifier resolution services associated with their data.

As a self-archiving service, researchers upload their datasets directly to DANS/EASY. Ingest processes are becoming more automated, especially for partners who contribute large numbers of datasets; data export services are being planned. Other planned technical services include linking datasets between DANS and Current Research Information Systems (CRIS) used to manage university publication and research records ("euroCRIS | Current Research Information Systems," 2017). Several application programming interfaces (API) to DANS databases and archives are under development. Some contributors to EASY provide searching access via their own websites, delivering datasets from EASY in the background.

*Acquiring and Curating Data*

DANS archivists solicit data for the collection by reading journals in their covered domains by attending conferences and holding workshops, and by contacting prospective contributors directly. They have developed a community that contributes datasets for stewardship and long-term access. The amount of labor required to process data depends on the condition of submitted datasets and the availability of relevant domain expertise. Some datasets arrive in standard formats that are easily ingested. Others require format migration, additional metadata, and various integrity checks. DANS staff take a comprehensive view that datasets are comprised of data, metadata, and persistent identifiers, whereas contributors tend to have a narrower perspective. DANSstaff5 mentioned that "it's sometimes hard to explain to people, to our consumers, or our contributors that I need to add something." DANSstaff2 provided a typology: "Dataset, in my perception, is the combination of adequate metadata in Dublin Core fields, possibly supplemented with specific metadata for a certain research specialization like language studies… The second element… is good documentation of the research project and its questions and its methodology and its problems perhaps. And the third element is of course an organized set of data files that are intelligible for somebody who finds them, that is organized in a good fashion, and that is in either accepted or preferred data formats."

Archivists reported that contributors often view self-archiving as simply depositing data "as is." Staff spend considerable labor training contributors in how to create the metadata necessary to make their data more useful to others. Some contributors were explicitly grateful for archivists' assistance. Contributor4 complemented the staff for migrating data from a proprietary format to a simpler format that is more widely used in the community. Contributor6 mentioned his great relief when a DANS archivist returned his dataset after finding personally identifiable data that were inadvertently left in the file. He cleaned the data and returned them to DANS for ingest.

DANS staff differed in their opinions of how to balance data acquisition and data curation. The self-archiving imperative justifies time spent on collection building, expecting that more data attracts more contributors and consumers, in a virtuous cycle. Conversely, others argued for more time spent on curation; by making datasets easier to find and use, the value of DANS services would increase. Some staff emphasized the responsibility of contributors to add metadata as part of the self-archiving processes, whereas others viewed metadata and curation as a staff function. Archivists consider themselves better at describing datasets in ways that others could find and use them. Curation activities include adding metadata, provenance, and documentation; migrating data to more standard formats; and other domain-specific improvements. Several mentioned the need to improve the search capabilities of EASY.

*Stewardship and Access to Data*

DANS staff balance current investments in dataset acquisition and curation with investments in long-term preservation and stewardship. Tradeoffs are many. New archaeological finds may have short and long-term value, which argues for placing high priority on curation and ingest. Oral histories, some of which are embargoed for years, may be less urgent, but important to preserve as cultural records. Dataset retrieval is unevenly distributed across the archive and patterns are difficult to anticipate, hence expected frequency of usage is not a viable metric.

DANS' operating principle is "Open if possible, protected where necessary" (Data Archiving and Networked Services, 2017). While the staff would generally prefer to release datasets openly under Creative Commons licenses, they recognize that they can acquire some important datasets only by allowing those contributors to maintain a degree of control over access. Locked files appear to get less usage due to the overhead of registering and requesting permission. DANS staff learn of access problems only when consumers contact them for assistance. Legal contracts that govern data deposit include mechanisms for control of locked data to default to DANS. Such policies are essential, otherwise datasets may remain locked indefinitely if the contributors do not respond, cannot be found, have left the institution, or are deceased.

# DISCUSSION

Our findings, drawn from weblogs, interviews, document analyses, and ethnography characterize the practices, policies, motivations, and concerns of stakeholders in DANS. Having summarized the findings by research question in the prior section, the discussion is organized thematically. Here we examine how knowledge infrastructures mediate access to data, who uses DANS, how and why they do so, characteristics and intersections of stakeholders, and the implications of these findings for sharing, reusing, and stewarding research data.

### How Knowledge Infrastructures Mediate Data Sharing and Reuse

Infrastructures are difficult to study because they are most visible when they break down and least visible when functioning well (Borgman, 2000; Edwards et al., 2013; Karasti & Blomberg, 2017; Star et al., 2003; Star & Ruhleder, 1996). DANS mediates access to data by providing human, technical, and policy infrastructure for their communities (Lee et al., 2006).

*Relationships between Stakeholders*

DANS technical architecture depends upon institutional relationships with partners in the Netherlands, the European Union, the U.S., and elsewhere. Most data are stored in national computing centers, for example, rather than in the DANS office complex in The Hague. For DANS data to be harvested by organizations such as OCLC, Google Scholar, and Europeana, they maintain organizational agreements and technical capabilities that allow datasets to be indexed and retrieved. In cases where datasets in DANS are delivered via APIs (application programming interfaces), the mediating role of DANS may not be apparent to consumers. The human infrastructure necessary for DANS to identify, acquire, curate, and sustain access to datasets is invisible to most users. These infrastructure activities require staff with high levels of expertise in technical standards, software engineering, law, and science policy. Whereas DANS contributors may have personal relationships with staff, consumers may encounter a staff member only when asking for assistance from the help desk or attending an information session.

Relationships between stakeholders are presented in Figure 2.

*Figure 2: Relationships between Stakeholders*

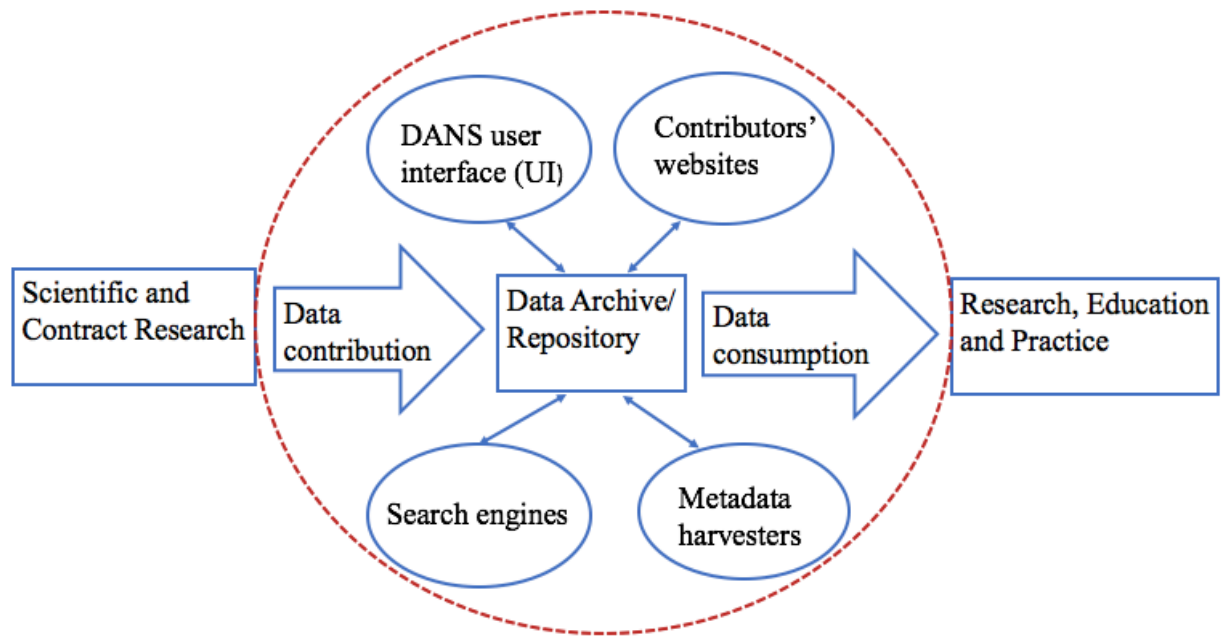

*Value Propositions and Metrics*

DANS, like any other public entity, must demonstrate "value for money." The value of data reuse is hard to evaluate in economic terms because uses are far downstream from data creation and ingest. Rare is the feedback loop to data creators, given the long provenance chains from data deposit to reusers who may mine and combine data from many sources. Policy makers and funders tend to place higher value on acquisition than on preservation, which is less visible and often more expensive. Datasets, once acquired, are stored, backed up, and migrated to new generations of technology. Even if data preservation agreements expire, managing those data and disposing of them requires money and effort.

Archaeology data constitute the largest portion of the archive, largest portion of acquisitions, and largest overall number of downloads (Doorn, 2017), but the most downloaded individual datasets are from the social sciences, such as census data from Statistics Netherlands (Akdag Salah et al., 2012; Doorn, 2017; Scharnhorst et al., 2012). The originating agencies do most of the processing, thus these datasets get high usage for relatively low investment in curation by DANS staff. These contrasts between investing in preservation and access highlight the complexity of DANS' mission.

DANS can claim a broad user base by counting registered users, uploads, searches, and downloads. To enumerate the community fully would require forcing users to register, to identify themselves uniquely, and for DANS to track user activities in detail. Such monitoring would likely run afoul of E.U. privacy regulations and create additional barriers to access ("General Data Protection Regulation (GDPR)," 2018).

Data citation is often mentioned as a way to evaluate digital data archives. Despite progress in assigning metadata Digital Object Identifiers to datasets, data citation is not yet common scholarly practice (Brase et al., 2014; CODATA-ICSTI Task Group on Data Citation Standards Practices, 2013; Data Archiving and Networked Services, 2010; Uhlir, 2012). DANS consumers rarely cite the data they acquire or cite DANS as a source. More problematic is that data contributors do not consistently cite their own datasets in DANS.

## Who Uses Digital Data Archives and Why?

Despite the best efforts of DANS and other digital data archives to serve their designated communities, these communities can be amorphous and evolve considerably over time. Archives with broad disciplinary responsibility and those who rely primarily on self-archiving by contributors are particularly challenged in identifying the scope and motivations of their communities. This study reveals the breadth of communities served by DANS, the diverse origins of datasets contributed, and the myriad uses to which those datasets may be put.

### *Scope of Datasets*

In DANS/EASY as elsewhere, notions of data are in the eye of the beholder (Borgman, 2015). Contributors of archaeology datasets made fine distinctions between excavation data, lists of finds, and reports on excavations. One archaeologist distinguished between boreholes and the multiple samples that might be drawn from each borehole. Several interviewees distinguished between digitized versions of analog records, such as excavation drawings, and born-digital data such as robotic measurements of field sites. Lines and polygons may be useful data in themselves; to become maps, these elements must be extracted with customized scripts and algorithms, often with proprietary software. In oral history, audio recordings of interviews, transcripts, summaries, and metadata each may be viewed as data. When consumers sought background information about a site, an excavation report in the form of a 90-page PDF was an adequate unit of data. In other cases, consumers preferred "raw data" in the most discrete units possible so that data could be aggregated in other ways. Plentiful content in Dutch, usually with English abstracts, makes DANS a rich resource, but it also limits the ability to provide standardized vocabularies, units of measurement, or common search features.

### *Uses and Reuses of Data*

The ability to create collections that can be mined and combined is a key driver of open data policies (Arzberger et al., 2004; Borgman, 2015; Boulton et al., 2015; Committee on Issues in the Transborder Flow of Scientific Data, National Research Council, 1997; Holdren, 2013). Whereas contributors often locked their data files to avoid mass downloading of datasets that might be reused without attribution, we encountered only a few cases in which data were being extracted as digital entities to be recombined with other data. Several consumers extracted data manually from DANS datasets, such as the local archaeology guide who made tables and images for the printed tour binder. Practitioners in the archaeology company mined excavation reports for data on sites of interest.

We did encounter similar contrasts in data reuse to those identified in studies of data practices in other domains. "Foreground uses" of data are cases where datasets are employed to create new knowledge products, whereas "background uses" are those in which datasets are used for general information about a site or a problem, such as calibration or history data (Pasquetto, 2018; Pasquetto, Borgman, & Wofford, in review; Wallis et al., 2013). Background uses of DANS datasets were most common in this study. The archaeology company started every investigation with a search of DANS/EASY to learn what is known about a site or region, for example. The local guide, the museum curator, and many others browsed DANS/EASY to learn what had been done before. Background uses of data are especially valuable at the early stages of research, but are particularly hard to identify, as they go unmentioned in research methods sections or in citation lists. One consumer used oral history archives for both background about projects and foreground, in which he mined datasets for linguistic structure.

### *Mediating Openness*

Digital data archives such as DANS/EASY are not simply publishing platforms in which contributors deposit data for anyone to use. They mediate open access to data in several ways. One way is by providing the infrastructure – human, technical, and institutional – that facilitates deposit, retrieval, and stewardship (Lee et al., 2006). Another way is by governing the rules of exchange between contributors and consumers. Whereas deposit with Creative Commons licenses would minimize the mediation required, that model would constrain DANS ability to acquire data from academic researchers. This community expressed greater willingness to submit data if they could maintain control over who has access to their data. By locking datasets, they can force potential consumers to register with DANS by name and to contact the contributor directly to request access. The access request process creates a side channel for contributors and consumers to negotiate access to datasets. If datasets are orphaned, with no response to access requests, control defaults to DANS. In the best cases, a fruitful conversation leads to selective sharing of appropriate datasets, and perhaps to collaboration. Because data are so difficult to interpret outside their original context, these personal relationships can be essential to reuse (Pasquetto et al., in review, 2017; Wallis et al., 2013).

### *Characteristics of Communities*

One goal of our study was to characterize the user community of DANS/EASY, especially with regard to the degree of overlap between contributors and consumers of data. The question of whether DANS' designated community (Consultative Committee for Space Data Systems, 2012) is one or several communities could not be answered from the weblogs. Our interviews suggest that the overlap between contributors and consumers of DANS/EASY data is minimal. These are largely distinct communities, more like book authors and readers than like a community of genomics researchers for which system input and output might be more equitable. Contributors tend to be researchers in universities, practitioners in archaeology companies, or institutions such as government statistics agencies. Consumers are a diverse array of individuals in universities, research institutes, museums, libraries, cultural institutions, and individuals who have avid avocations in areas where DANS/EASY has content. They may search infrequently, but rely on the system as an essential resource.

# CONCLUSIONS

By studying the uses and users of DANS as an exemplar of a digital data archive, we reveal the function of the archive in knowledge infrastructures and ways in which the archive mediates information exchange in communities. Data archives as institutions, and the archivists who work directly with contributors and consumers of data, are active participants in the communities they serve. Data archives are not passive repositories, nor are they simply databases of content.

### Communities and Craft

Automation is facilitating more archival procedures, such as batch ingest of files and APIs for submission and retrieval, but much of the labor associated with contributing data to archives remains craft work. With the exception of institutions that submit large numbers of files on a regular basis, such as government statistics agencies and archaeology contractors, most contributors appear to submit files only once every year or two. Search and retrieval have similar patterns, with a few large consumers and many occasional users. Several of the consumers interviewed mentioned that they had refreshed their familiarity with DANS/EASY prior to meeting with us. The actual number of DANS/EASY users cannot be known, as registration cannot be required for privacy reasons. The patterns of use revealed in our interviews are similar to the long-tail distributions evident in other studies of information-seeking behavior (Case, 2006).

Given the long-tail characteristics of both contributors and consumers, and the deliberately diverse array of data that DANS/EASY collects, little of the archival work is subject to standardization or automation. Contributors who submit a dataset once every year or two, or maybe once in a career, need assistance in structuring and documenting their files for submission. Archivists need to verify metadata, documentation, and data integrity to ensure that datasets meet minimum standards for ingest. "Self-archiving" is accomplished with professional assistance, lest DANS, or any other archive, be littered with unusable data files. Metadata standards and classification can assure some basic level of discovery, but standardizing formats and vocabularies across content that spans government statistics, archaeological digs, oral histories, and biological records is nigh unto impossible. More investment in metadata, documentation, and retrieval tools would enhance discovery, but tradeoffs in these labor-intensive investments are necessary. Human infrastructure is expensive, but essential, for sustaining access to research data (Borgman, 2015; Lee et al., 2006).

Only the staff of the archaeology company were frequent searchers of DANS, starting each new investigation with a search for what is already known about a site. Some of the contributors claimed not to search DANS at all, instead viewing DANS as an institution to steward and provide access to the content they contributed.

## Mediation

DANS, as a digital data archive, mediates data sharing and reuse in a variety of ways. The most obvious mediation is between contributors and consumers. Those who create and seek any given dataset are unlikely to find each other on their own. Rather, the creators share their data by contributing them to DANS and the consumers find data by searching DANS. Less obvious are the temporal aspects of these exchange relationships. The time of contribution and consumption may be days – or decades – apart. DANS thus mediates between communities whose membership changes continuously.

Data archives' mediation between contributors and consumers requires several steps, each of which may involve multiple parties and processes. A data creator in a university may submit a dataset to library staff who serve as a "back office" to DANS. The library staff will do some of the processing, such as verifying formats, metadata, file structures, privacy protection, and data integrity. Some of this work may be left to DANS staff. Contributors may do some of this work themselves, especially in high throughput environments such as the archaeology companies. Other contributors wish to deposit their data "as is," leaving more work for the archivists to bring datasets up to minimal standards – which may require multiple interactions with the contributor and delays in processing.

DANS mediates between contributors and consumers in the cases of locked files. This is also a multilayered relationship. At the highest level is setting policy for the degrees of control that a contributor can retain over files, once ingested. For many individuals, the more control they can retain, the more willing they are to contribute their data. Conversely, the lesser the degree of contributor control, the easier it is for archives to provide access to data and for consumers to reuse them. DANS sets policy for locking and unlocking files, then leaves negotiation to the parties involved. Only when those relationships break down does DANS intervene; they reserve the right to claim control over the datasets when contributors do not satisfy their contractual obligations.

## Knowledge Infrastructures

The knowledge infrastructure in which DANS functions is broad, encompassing many public and private stakeholders around the world, few of which existed when the DANS collections were formed more than five decades ago. These include providers of search engines, library networks, cultural heritage portals, and websites that harvest and serve DANS data. More narrowly, DANS is a node in the knowledge infrastructures of the communities whose data they acquire and of those who consume their data. DANS deploys technology, sets policy, writes contracts, and stewards the datasets in their care. They also build communities by soliciting datasets, training, and outreach. As these communities evolve over long periods of time, the digital data archive provides continuity, connecting generations past, present, and future.

Digital data archives such as DANS are investments that keep on giving over long periods of time. They are expensive, labor-intensive, hard to measure, and evolving. They take many years to build but can degrade quickly through lack of continuous investment. They must constantly prove their value to their communities, as must any organization. However, as an information institution, their most important communities – the generations of the future – have not yet been born. The value of digital data archives in data sharing and reuse can be evaluated only by taking a very long view.

# ACKNOWLEDGEMENTS

We gratefully acknowledge the support of KNAW, the Netherlands Academy of Arts and Sciences, for Visiting Scholar appointments at DANS of Christine L. Borgman, Herbert Van de Sompel, and Andrew Treloar. Peter Doorn, Director of DANS, graciously opened the doors, physically and digitally, for these investigations, and provided Andrea Scharnhost's time for the project. Additional support for conducting and analyzing interviews is provided by grant # 201514001, from the Alfred P. Sloan Foundation to UCLA, Christine L. Borgman, Principal Investigator. The EC funded FP7 project Impact-EV has provided support for data mining, analytics and methods. COST TD1210 also supported dissemination and meetings. Interviews with human subjects in the Netherlands are covered under UCLA IRB # 15-001291. Most interviews were conducted by Christine L. Borgman and Andrea Scharnhorst; Ashley Sands and Milena Golshan conducted the remainder. Data analysis was conducted by Milena Golshan and Christine L. Borgman. Weblog analysis was conducted by Henk van den Berg.

Andrew Treloar and Herbert Van de Sompel contributed to the initial design of the study. Sally Wyatt provided expert guidance throughout the study and commentary on interim drafts. Willeke Wendrich and Deidre Whitmore provided expertise on archaeological uses of data. Ashley Sands, Peter Darch, Irene Pasquetto, and Bernadette Boscoe also offered guidance in project design, data analysis, and comments on drafts. Most of all, we acknowledge the generosity of DANS staff for their thoughtful discussions of work practices, and of DANS/EASY contributors and consumers who welcomed us into their offices and homes or traveled to meet us elsewhere in the Netherlands.

# REFERENCES

Advanced Research Infrastructure for Archaeological Dataset Networking. (2012). KNAW-DANS | ARIADNE. Retrieved July 10, 2018, from http://www.ariadne-infrastructure.eu/About/Consortium/KNAW-DANS

Akdag Salah, A. A., Scharnhorst, A., ten Bosch, O., Doorn, P. K., Manovich, L., Salah, A. A., & Chow, J. (2012). Significance of Visual Interfaces in Institutional and User-generated Databases with Category Structures. In *Proceedings of the Second International ACM Workshop on Personalized Access to Cultural Heritage* (pp. 7–10). New York, NY, USA: ACM. https://doi.org/10.1145/2390867.2390870

Arbuckle, B. S., Kansa, S. W., Kansa, E. C., Orton, D., Çakırlar, C., Gourichon, L., … Würtenberger, D. (2014). Data Sharing Reveals Complexity in the Westward Spread of Domestic Animals across Neolithic Turkey. *PLoS ONE*, *9*(6), e99845. https://doi.org/10.1371/journal.pone.0099845

Arzberger, P., Schroeder, P., Beaulieu, A., Bowker, G. C., Casey, K., Laaksonen, L., … Wouters, P. (2004). An International Framework to Promote Access to Data. *Science*, *303*(5665), 1777–1778. https://doi.org/10.1126/science.1095958

Borgman, C. L. (2000). *From Gutenberg to the Global Information Infrastructure: Access to Information in the Networked World*. Cambridge, MA: MIT Press.

Borgman, C. L. (2015). *Big data, little data, no data: Scholarship in the networked world*. Cambridge, MA: MIT Press.

Borgman, C. L. (2018). Open Data, Grey Data, and Stewardship: Universities at the Privacy Frontier. *Berkeley Technology Law Journal*, *33*(2), 365–412. https://doi.org/10.15779/Z38B56D489

Borgman, C. L., Darch, P. T., Sands, A. E., & Golshan, M. S. (2016). The durability and fragility of knowledge infrastructures: Lessons learned from astronomy. In *Proceedings of the Association for Information Science and Technology* (Vol. 53, pp. 1–10). Copenhagen, Denmark. Retrieved from http://dx.doi.org/10.1002/pra2.2016.14505301057

Borgman, C. L., Darch, P. T., Sands, A. E., Pasquetto, I. V., Golshan, M. S., Wallis, J. C., & Traweek, S. (2015). Knowledge infrastructures in science: Data, diversity, and digital libraries. *International Journal on Digital Libraries*, *16*(3–4), 207–227. https://doi.org/10.1007/s00799-015-0157-z

Borgman, C. L., Darch, P. T., Sands, A. E., Wallis, J. C., & Traweek, S. (2014). The Ups and Downs of Knowledge Infrastructures in Science: Implications for Data Management. In *2014 IEEE/ACM Joint Conference on Digital Libraries (JCDL)* (pp. 257–266). London: IEEE Computer Society. https://doi.org/10.1109/JCDL.2014.6970177

Borgman, C. L., Golshan, M. S., Sands, A. E., Wallis, J. C., Cummings, R. L., Darch, P. T., & Randles, B. M. (2016). Data Management in the Long Tail: Science, Software, and Service. *International Journal of Digital Curation*, *11*(1), 128–149. https://doi.org/10.2218/ijdc.v11i1.428

Borgman, C. L., Hirsh, S. G., & Hiller, J. (1996). Rethinking online monitoring methods for information retrieval systems: From search product to search process. *Journal of the American Society for Information Science*, *47*(7), 568–583. https://doi.org/10.1002/(SICI)1097-4571(199607)47:7<568::AID-ASI8>3.0.CO;2-S

Borgman, C. L., Van de Sompel, H., Scharnhorst, A., van den Berg, H., & Treloar, A. (2015). Who Uses the Digital Data Archive? An Exploratory Study of DANS. In *Proceedings of the Association for Information Science and Technology* (Vol. 52). St. Louis, MO. https://doi.org/10.1002/pra2.2015.145052010096

Boulton, G., Babini, D., Hodson, S., Li, J., Marwala, T., Musoke, M. G. N., … Wyatt, S. (2015). *Open Data in a Big Data World: An International Accord* (Outcome of Science International 2015 meeting). ICSU, IAP, ISSC, TWAS. Retrieved from https://twas.org/sites/default/files/open-data-in-big-data-world_short_en.pdf

Bowker, G. C. (2005). *Memory Practices in the Sciences*. Cambridge, Mass.: MIT Press.

Brase, J., Socha, Y., Callaghan, S., Borgman, C. L., Uhlir, P. F., & Carroll, B. (2014). Data Citation: Principles and Practice. In J. M. Ray, *Research Data Management: Practical Strategies for Information Professionals*. West Lafayette: Purdue University Press.

Case, D. O. (2006). *Looking for Information: A Survey of Research on Information Seeking, Needs, and Behavior* (2nd ed.). San Diego: Academic Press.

CODATA-ICSTI Task Group on Data Citation Standards Practices. (2013). Out of Cite, Out of Mind: The Current State of Practice, Policy, and Technology for the Citation of Data. *Data Science Journal*, *12*, CIDCR1–CIDCR75. https://doi.org/10.2481/dsj.OSOM13-043

Committee on Issues in the Transborder Flow of Scientific Data, National Research Council. (1997). *Bits of Power: Issues in Global Access to Scientific Data*. National Academies Press. Retrieved from http://www.nap.edu/openbook.php?record_id=5504

Consultative Committee for Space Data Systems. (2012). *Reference model for an Open Archival Information System (OAIS)* (Recommendation for space data system practices No. CCSDS 650.0-M-2 Magenta Book). Washington, D.C. Retrieved from https://public.ccsds.org/pubs/650x0m2.pdf

Council of Europe. (1992). European Convention on the Protection of the Archaeological Heritage: Valletta Treaty No. 143. Retrieved May 19, 2017, from https://www.coe.int/en/web/conventions/full-list/-/conventions/treaty/143

Curty, R. G., Crowston, K., Specht, A., Grant, B. W., & Dalton, E. D. (2017). Attitudes and norms affecting scientists' data reuse. *PLOS ONE*, *12*(12), e0189288. https://doi.org/10.1371/journal.pone.0189288

Darch, P. T., & Borgman, C. L. (2016). Ship space to database: emerging infrastructures for studies of the deep subseafloor biosphere. *PeerJ Computer Science*, *2*, e97. https://doi.org/10.7717/peerj-cs.97

Darch, P. T., Borgman, C. L., Traweek, S., Cummings, R. L., Wallis, J. C., & Sands, A. E. (2015). What lies beneath?: Knowledge infrastructures in the subseafloor biosphere and beyond. *International Journal on Digital Libraries*, *16*(1), 61–77. https://doi.org/10.1007/s00799-015-0137-3

Data Archiving and Networked Services. (2010). *The first five years of Data Archiving and Networked Services: Self-assessment DANS 2005 - 2010* (p. 53). The Hague: DANS. Retrieved from https://www.knaw.nl/shared/resources/actueel/bestanden/DANS_Self-assessment_report_2010.pdf

Data Archiving and Networked Services. (2017). DANS: Organisation and policy. Retrieved July 12, 2017, from https://dans.knaw.nl/en/about/organisation-and-policy

Digital Preservation Coalition. (2018). Home page. Retrieved July 12, 2018, from https://www.dpconline.org/

Dillo, I., & Doorn, P. K. (2011). *The Dutch data landscape in 32 interviews and a survey*. Retrieved from http://depot.knaw.nl/10090/1/The_Dutch_Datalandscape_DEF.pdf

Dillo, I., & Doorn, P. K. (2014). The Front Office–Back Office Model: Supporting Research Data Management in the Netherlands. *International Journal of Digital Curation*, *9*(2), 39–46. https://doi.org/10.2218/ijdc.v9i2.333

Directory of Open Access Repositories - SHERPA Services. (2018). Retrieved June 29, 2018, from http://v2.sherpa.ac.uk/opendoar/

Doorn, P. K. (2017). *Datametric Analysis of DANS Data Archive* (No. version 1.0). DANS.

Doorn, P. K., & Tjalsma, H. (2007). Introduction: archiving research data. *Archival Science*, *7*(1), 1–20. https://doi.org/10.1007/s10502-007-9054-6

EASY: Published datasets. (2016). Retrieved September 19, 2016, from https://easy.dans.knaw.nl/ui/browse

Edmunds, R., L'Hours, H., Rickards, L., Trilsbeek, P., & Vardigan, M. (2016). Core Trustworthy Data Repositories Requirements. *Zenodo*. https://doi.org/10.5281/zenodo.168411

Edwards, P. N. (2010). *A Vast Machine: Computer Models, Climate Data, and the Politics of Global Warming*. Cambridge, MA: The MIT Press.

Edwards, P. N., Jackson, S. J., Chalmers, M. K., Bowker, G. C., Borgman, C. L., Ribes, D., … Calvert, S. (2013). *Knowledge infrastructures: Intellectual frameworks and research challenges* (p. 40). Ann Arbor, MI: University of Michigan.

Eschenfelder, K. R., & Johnson, A. (2014). Managing the data commons: Controlled sharing of scholarly data. *Journal of the Association for Information Science and Technology*, *65*(9), 1757–1774. https://doi.org/10.1002/asi.23086

EUDAT Collaborative Data Infrastructure. (2016, November 4). Partners | EUDAT. Retrieved July 10, 2018, from https://www.eudat.eu/eudat-cdi/partners

euroCRIS | Current Research Information Systems. (2017). Retrieved July 28, 2017, from http://www.eurocris.org/

European Holocaust Research Infrastructure. (2015, June 16). EHRI Partners [Text]. Retrieved July 10, 2018, from https://www.ehri-project.eu/ehri-partners

European Open Science Cloud. (2018, January 3). EOSC-hub: integrated services for the European Open Science Cloud. Retrieved July 10, 2018, from https://eosc-hub.eu/news/eosc-hub-integrated-services-european-open-science-cloud

Faniel, I. M., Barrera-Gomez, J., Kriesberg, A., & Yakel, E. (2013). A comparative study of data reuse among quantitative social scientists and archaeologists. In *iConference 2013 Proceedings* (pp. 797–800). https://doi.org/10.9776/13391

Faniel, I. M., Kansa, E. C., Kansa, S. W., Barrera-Gomez, J., & Yakel, E. (2013). The Challenges of Digging Data: A Study of Context in Archaeological Data Reuse. In *Proceedings of the 13th ACM/IEEE-CS Joint Conference on Digital Libraries* (pp. 295–304). New York, NY, USA: ACM. https://doi.org/10.1145/2467696.2467712

Faniel, I. M., & Yakel, E. (2017). Practices Do Not Make Perfect: Disciplinary Data Sharing and Reuse Practices and Their Implications for Repository Data Curation. In L. R. Johnston (Ed.), *Curating Research Data, Volume One: Practical Strategies for Your Digital Repository* (pp. 103–126). Chicago, Illinois: Association of College and Research Libraries. Retrieved from http://www.oclc.org/research/publications/2017/practices-do-not-make-perfect.html

Force11. (2018). About Force11. Retrieved June 26, 2015, from https://www.force11.org/about

Frank, R. D., Yakel, E., & Faniel, I. M. (2015). Destruction/reconstruction: preservation of archaeological and zoological research data. *Archival Science*, *15*(2), 141–167. https://doi.org/10.1007/s10502-014-9238-9

General Data Protection Regulation (GDPR). (2018). Retrieved July 6, 2018, from https://gdpr-info.eu/

Gregory, K., Cousijn, H., Groth, P., Scharnhorst, A., & Wyatt, S. (2018). Understanding Data Retrieval Practices: A Social Informatics Perspective. *ArXiv:1801.04971 [Cs]*. Retrieved from http://arxiv.org/abs/1801.04971

Holdren, J. P. (2013, February 22). Increasing Access to the Results of Federally Funded Scientific Research. Executive Office of the President, Office of Science and Technology Policy. Retrieved from https://www.usaid.gov/sites/default/files/documents/1865/NW2-CCBY-HO2-Public_Access_Memo_2013.pdf

International Council of Scientific Unions. (2018). World Data System: Trusted Data Services for Global Science. Retrieved July 3, 2017, from https://www.icsu-wds.org/

Kansa, E. C. (2012). Openness and archaeology's information ecosystem. *World Archaeology*, *44*(4). https://doi.org/10.1080/00438243.2012.737575

Kansa, E. C., & Kansa, S. W. (2011). Toward a Do-It-Yourself Cyberinfrastructure: Open Data, Incentives, and Reducing Costs and Complexities of Data Sharing. In *Archaeology 2.0: New Approaches to Communication and Collaboration* (pp. 57–92). Cotsen Digital Archaeology series. Retrieved from http://escholarship.org/uc/item/1r6137tb

Kansa, E. C., & Kansa, S. W. (2013). We All Know That a 14 Is a Sheep: Data Publication and Professionalism in Archaeological Communication. *Journal of Eastern Mediterranean Archaeology and Heritage Studies*, *1*(1), 88–97. Retrieved from https://muse.jhu.edu/article/501744

Kansa, E. C., Kansa, S. W., & Arbuckle, B. (2014). Publishing and Pushing: Mixing Models for Communicating Research Data in Archaeology. *International Journal of Digital Curation*, *9*(1), 57–70. https://doi.org/10.2218/ijdc.v9i1.301

Karasti, H., & Blomberg, J. (2017). Studying Infrastructuring Ethnographically. *Computer Supported Cooperative Work (CSCW)*. https://doi.org/10.1007/s10606-017-9296-7

Latour, B. (1987). *Science in action: How to follow scientists and engineers through society*. Cambridge, MA: Harvard University Press.

Latour, B., & Woolgar, S. (1986). *Laboratory Life: The Construction of Scientific Facts* (2nd ed.). Princeton, N.J.: Princeton University Press.

Lave, J., & Wenger, E. (1991). *Situated Learning: Legitimate Peripheral Participation*. Cambridge, UK: Cambridge University Press.

Lee, C. P., Dourish, P., & Mark, G. (2006). The Human Infrastructure of Cyberinfrastructure. In *Proceedings of the 2006 20th Anniversary Conference on Computer Supported Cooperative Work* (pp. 483–492). New York, NY, USA: ACM. https://doi.org/10.1145/1180875.1180950

Mayernik, M. S. (2011). *Metadata Realities for Cyberinfrastructure: Data Authors as Metadata Creators* (PhD Dissertation). UCLA, Los Angeles, CA. Retrieved from http://dx.doi.org/10.2139/ssrn.2042653

Mayernik, M. S. (2016). Research data and metadata curation as institutional issues. *Journal of the Association for Information Science and Technology*, *67*(4), 973–993. https://doi.org/10.1002/asi.23425

Mayernik, M. S., Wallis, J. C., & Borgman, C. L. (2013). Unearthing the Infrastructure: Humans and Sensors in Field-Based Research. *Computer Supported Cooperative Work*, *22*(1), 65–101. https://doi.org/10.1007/s10606-012-9178-y

Mayernik, M. S., Wallis, J. C., Pepe, A., & Borgman, C. L. (2008). Whose data do you trust? Integrity issues in the preservation of scientific data. In *Proceedings of iConference 2008: iFutures: Systems, Selves, Society*. Los Angeles, CA. http://hdl.handle.net/2142/15119

Mientjes, A. C. (2015). Archeologische Begeleiding, Protocol opgraven, Kruittoren 17 - 25, Tholen, Gemeente Tholen. *(SOB Research)*, *DANS*. https://doi.org/10.17026/dans-23t-32e4

National Science Board (U.S.). (2005). *Long-Lived Digital Data Collections: Enabling Research and Education in the 21st Century* (No. US NSF-NSB-05-40). Arlington, Virginia: National Science Foundation.

Palmer, C. L. (2005). Scholarly work and the shaping of digital access. *Journal of the American Society for Information Science and Technology*, *56*(11), 1140–1153. https://doi.org/10.1002/asi.20204

Pasquetto, I. V. (2018). *From Open Data to Knowledge Production: Biomedical Data Sharing and Unpredictable Data Reuses* (Ph.D. Dissertation). UCLA, Los Angeles, CA.

Pasquetto, I. V., Borgman, C. L., & Wofford, M. F. (in review). The Who, What, When, and Why of Reusing Data in Scientific Practice. *Harvard Data Science Review*.

Pasquetto, I. V., Randles, B. M., & Borgman, C. L. (2017). On the Reuse of Scientific Data. *Data Science Journal*, *16*. https://doi.org/10.5334/dsj-2017-008

Pasquetto, I. V., Sands, A. E., & Borgman, C. L. (2015). Exploring openness in data and science: What is "open," to whom, when, and why? In *Proceedings of the Association for Information Science and Technology* (Vol. 52, pp. 1–2). St. Louis, MO. https://doi.org/10.1002/pra2.2015.1450520100141

Piwowar, H. A. (2011). Who Shares? Who Doesn't? Factors Associated with Openly Archiving Raw Research Data. *PLoS ONE*, *6*(7), e18657. https://doi.org/10.1371/journal.pone.0018657

Regents of the University of Michigan. (2016). ICPSR - Inter-university Consortium for Political and Social Research. Retrieved April 2, 2013, from http://www.icpsr.umich.edu/icpsrweb/ICPSR/

Reijnhoudt, L., Stamper, M. J., Börner, K., Baars, C., & Scharnhorst, A. (2012). NARCIS: Network of Experts and Knowledge Organizations in the Netherlands. Map. › KNAW Research Portal. DANS-KNAW. Retrieved from http://hdl.handle.net/20.500.11755/88be28c1-4293-42e8-85b4-088433c03ae7

Research Data Alliance. (2018a). About RDA. Retrieved June 26, 2015, from https://rd-alliance.org/about.html

Research Data Alliance. (2018b). RDA: Research Data Alliance [Home page]. Retrieved September 30, 2013, from https://rd-alliance.org

Sands, A. E. (2017). *Managing Astronomy Research Data: Data Practices in the Sloan Digital Sky Survey and Large Synoptic Survey Telescope Projects* (Ph.D. Dissertation). UCLA, Los Angeles, CA. Retrieved from http://escholarship.org/uc/item/80p1w0pm

Scharnhorst, A., Bosch, O. ten, & Doorn, P. K. (2012). Looking at a digital research data archive - Visual interfaces to EASY. *ArXiv:1204.3200 [Physics]*. Retrieved from http://arxiv.org/abs/1204.3200

Science Europe. (2018). Science Europe. Retrieved July 11, 2018, from https://www.scienceeurope.org/

Shankar, K., Eschenfelder, K. R., & Downey, G. (2016). Studying the History of Social Science Data Archives as Knowledge Infrastructure. *Science & Technology Studies*, *29*(2). Retrieved from http://ojs.tsv.fi/index.php/sts/article/view/55691

Star, S. L., Bowker, G. C., & Neumann, L. J. (2003). Transparency beyond the Individual Level of Scale: Convergence between Information Artifacts and Communities of Practice. In A. Bishop, N. A. Van House, & B. P. Buttenfield (Eds.), *Digital library use: Social practice in design and evaluation* (pp. 241–270). Cambridge Mass.: MIT Press.

Star, S. L., & Ruhleder, K. (1996). Steps Toward an Ecology of Infrastructure: Design and Access for Large Information Spaces. *Information Systems Research*, *7*(1), 111–134. https://doi.org/10.1287/isre.7.1.111

Steinmetz Archive: Dutch Social Science Data Archive. (1989). *Historical Social Research / Historische Sozialforschung*, *14*(1 (49)), 118–121. Retrieved from http://www.jstor.org/stable/20754372

Tenopir, C., Dalton, E. D., Allard, S., Frame, M., Pjesivac, I., Birch, B., … Dorsett, K. (2015). Changes in Data Sharing and Data Reuse Practices and Perceptions among Scientists Worldwide. *PLOS ONE*, *10*(8), e0134826. https://doi.org/10.1371/journal.pone.0134826

Tenopir, C., Palmer, C. L., Metzer, L., van der Hoeven, J., & Malone, J. (2011). Sharing data: Practices, barriers, and incentives. *Proceedings of the American Society for Information Science and Technology*, *48*(1), 1–4. https://doi.org/10.1002/meet.2011.14504801026

Tsoukala, V., Angelaki, M., Kalaitzi, V., Wessels, B., Price, L., Taylor, M. J., … Wadhwa, K. (2015). *Policy RECommendations for Open Access to Research Data in Europe: RECODE Project*. Retrieved from http://recodeproject.eu/wp-content/uploads/2015/02/RECODE-D5.1-POLICY-RECOMMENDATIONS-_FINAL.pdf

UCLA Center for Knowledge Infrastructures: Home. (2018). Retrieved from https://knowledgeinfrastructures.gseis.ucla.edu/

Uhlir, P. F. (Ed.). (2012). *For Attribution -- Developing Data Attribution and Citation Practices and Standards: Summary of an International Workshop*. Washington, D.C.: The National Academies Press.

Wallis, J. C., Rolando, E., & Borgman, C. L. (2013). If We Share Data, Will Anyone Use Them? Data Sharing and Reuse in the Long Tail of Science and Technology. *PLOS ONE*, *8*(7), e67332. https://doi.org/10.1371/journal.pone.0067332

Weber, N. M., Baker, K. S., Thomer, A. K., Chao, T. C., & Palmer, C. L. (2012). Value and context in data use: Domain analysis revisited. *Proceedings of the American Society for Information Science and Technology*, *49*(1), 1–10. https://doi.org/10.1002/meet.14504901168

Wenger, E. (1998). *Communities of practice : learning, meaning, and identity*. Cambridge: Cambridge University Press.

Wilkinson, M. D., Dumontier, M., Aalbersberg, Ij. J., Appleton, G., Axton, M., Baak, A., … Mons, B. (2016). The FAIR Guiding Principles for scientific data management and stewardship. *Scientific Data*, *3*, 160018. Retrieved from http://dx.doi.org/10.1038/sdata.2016.18

Yakel, E., Faniel, I. M., Kriesberg, A., & Yoon, A. (2013). Trust in Digital Repositories. *International Journal of Digital Curation*, *8*(1), 143–156. https://doi.org/10.2218/ijdc.v8i1.251

Zimmerman, A. S. (2008). New Knowledge from Old Data: The Role of Standards in the Sharing and Reuse of Ecological Data. *Science, Technology & Human Values*, *33*(5), 631–652. https://doi.org/10.1177/0162243907306704